\documentclass[showpacs,preprintnumbers,amsmath,amssymb,twocolumn,superscriptaddress,prb]{revtex4}
\usepackage{graphicx}
\usepackage{amsfonts}
\usepackage{amsmath, amsthm, amssymb}
\usepackage{dsfont}
\usepackage{times}

\newcommand{\ve}[1]{\mathbf{#1}}
\newcommand{\vk}{\ve{k}} 

\newcommand{\e}[1]{\mathrm{e}^{#1}}

\newcommand{\eg}{\textit{e.g. }}
\newcommand{\etal}{\emph{et al.}}
\def\i{\mathrm{i}}

\begin{document}
\title[Strongly spin-polarized current generated in Zeeman-split unconventional superconductors]{Strongly spin-polarized current 
generated in a Zeeman-split unconventional superconductors}
\author{Jacob Linder}
\affiliation{Department of Physics, Norwegian University of
Science and Technology, N-7491 Trondheim, Norway}
\author{Takehito Yokoyama}
\affiliation{Department of Applied Physics, Nagoya University, Nagoya, 464-8603, Japan}
\author{Yukio Tanaka}
\affiliation{Department of Applied Physics, Nagoya University, Nagoya, 464-8603, Japan}
\author{Asle Sudb{\o}}
\affiliation{Department of Physics, Norwegian University of
Science and Technology, N-7491 Trondheim, Norway}

\date{Received \today}
\begin{abstract}
We consider a thin-film normal metal/superconductor junction in the presence of an externally applied in-plane magnetic field for several 
symmetries of the superconducting order parameter. For $p$-wave superconductors, a strongly spin-polarized current emerges due to an interplay between the nodal structure of the superconducting order parameter, the 
existence or non-existence of zero-energy surface states, and the Zeeman-splitting of the bands which form superconductivity. Thus, 
the polarization depends strongly on the orbital symmetry of the superconducting state. Our findings suggest a  mechanism 
for obtaining fully spin-polarized currents crucially involving zero-energy surface states, not present in $s$-wave superconductors. 

\end{abstract}

\maketitle
\section{Introduction}
In recent years, spintronics \cite{zutic,Tserkovnyak,Maekawa,Awschalom} has grown enormously as a research field, based on the idea 
that the electron spin may form a centerpiece in future technological applications. The main issues in this field are: \textit{i)} 
how may one obtain and manipulate the spin-polarization of an electrical current, and \textit{ii)} how may the spin-polarization 
of an electrical current be detected? 
\par
Concerning the first issue, suggestions so far (see Ref. \cite{zutic} and references therein) have mostly revolved around the use 
of semiconducting materials. These materials have the potential of offering some control over the spin injection properties via the 
coupling between the spin-degree of freedom and the electrons orbital motion. This coupling originates with the spin-orbit coupling 
that is present in such materials. Concerning the second question, detection of a spin-current has been proposed in the form of spin 
accumulation \cite{accumulate,Kato,Wunderlich} and that a spin-current should generate electrical fields \cite{detect}. 
\par
In the search for functionalities utilizing ideas involving the spin of the electrons (spintronics), a subfield known 
as \textit{superspintronics}.
The idea is to combine the useful properties of superconductors with 
spin generation and manipulation \cite{brataas, eremin, linder, eschrig, dani}. Most known superconductors have a spin-singlet symmetry, which 
means that the Cooper pair does not carry any net spin. For such superconductors, one relies mostly on strong magnetic sources 
such as half-metallic ferromagnets in comcomitance with superconductors  for obtaining strongly spin-polarized currents. 
\par
Recently, however, it was suggested in Ref. \cite{taddei} that a thin-film $s$-wave superconductor subjected to an in-plane magnetic 
field may serve to strongly spin-polarize electrical currents in the tunneling limit. This actually follows from results obtained by Meservey and Tedrow \cite{meservey} who performed experiments in spin-polarized electron tunneling in thin-film $s$-wave superconductors subjected to an in-plane magnetic field. In Ref. \cite{dani}, a proposal for an absolute spin-valve effect was put forth without assuming a thin-film structure of the spin-singlet superconductor. On the other hand, there now exist several 
superconductors exhibiting spin-triplet superconductivity  \cite{maeno,Tou_1998,Lee_2002,Harada_2007}, and these systems 
are less antagonistic towards an applied magnetic fields than what their spin-singlet counterparts are. 
\par
An intriguing situation may arise when an in-plane magnetic field is applied to a thin-film superconductor. If the thickness $d$ 
of the superconducting film satisfies $d\ll\lambda$, where $\lambda$ is the magnetic penetration depth, the field penetrates the 
superconducting film homogeneously and induces a Zeeman-splitting of the bands. Experiments on such structures have clearly 
revealed a spin-split density of states in the superconductor \cite{meservey}, and the problem was recently re-examined in Ref.~\onlinecite{taddei}.
\par
\begin{figure}[h!]
\centering
\resizebox{0.46\textwidth}{!}{
\includegraphics{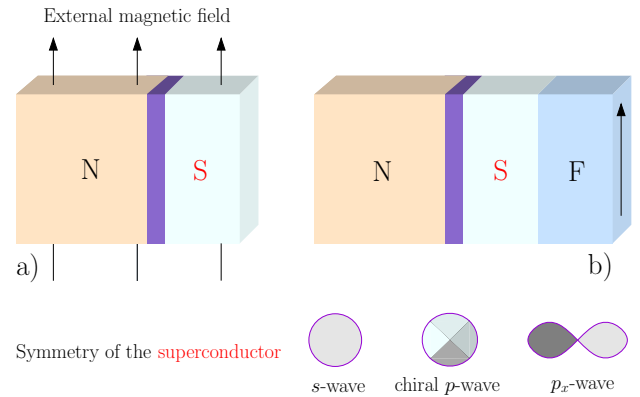}}
\caption{(Color online) The experimental setup proposed in this paper. In a), an in-plane magnetic field is applied to the 
junction. In b), an exchange field is induced by a ferromagnetic film in close proximity. We will consider three different 
symmetries of the superconducting state, as shown above. In the chiral p-wave case, the gap has an intrinsic complex angular
dependence and a constant magnitude. The system we consider is similar to that of Ref.~\onlinecite{taddei}.}
\label{fig:model}
\end{figure}
A natural question arises in the context of \textit{Zeeman-split superconductors}: what is the effect of the orbital symmetry of 
the superconducting order parameter on the polarization of the electrical current? In this work, we show that the 
\textit{orbital symmetry} of the superconducting state strongly influences the \textit{spin-polarization} of the electrical current. 
We consider three different orbital symmetries for the superconductor, and show how the polarization properties of the current 
differ greatly in each case, even though the spin structure is similar for each superconducting state. It follows from our results 
that the polarization properties of the current may be used not only as a tool for obtaining information about the orbital symmetry 
of the superconducting state, but that {\it the spin-polarization may be controlled efficiently by a bias voltage due to an interplay 
between superconductivity and magnetism}.\cite{meservey,dani,taddei} The physics is that the Zeeman-splitting of the bands leads to an onset of electrical 
currents of  majority and minority spin species at distinct bias voltages. This phenomenon combines with a subtle enhancement of 
the conductance in a given spin channel which is determined by a resonance condition that sensitively depends on the orbital
symmetry of the superconductiong order parameter. In this context, it will be shown that zero-energy surface states play a 
crucial role. We now proceed to present our results in detail.

\section{Theory}
When an in-plane magnetic field is applied to a thin-film superconductor, there is an upper critical field associated with a first 
order phase transition from the superconducting to paramagnetic state. The upper critical field may be determined by considering 
the argument of Ref. \cite{clogston}, and essentially consists of balancing the free energies in the paramagnetic and superconducting 
state. Extending this argument to anisotropic superconductors, we find that the critical value for the exchange splitting $h$ in the 
superconductor reads
\begin{align}
h_c = \Delta_0 \sqrt{\langle |g_\vk|^2\rangle/2},\; \Delta_\vk = \Delta_0 g_\vk.
\end{align}
Here, $\Delta_\vk$ is the gap function with its magnitude $\Delta_0$ and $\langle \ldots\rangle$ denotes angular average over the Fermi 
surface. We have $g_\vk=1$ for $s$-wave superconductors, $g_\vk=\e{\i\theta}$ for chiral $p$-wave superconductors, and 
$g_\vk = \cos\theta$ for $p_x$-wave superconductors, where $\theta$ is the azimuthal angle. Considering $T=0$, a 
self-consistent solution of the order parameter reveals that the value of the gap is constant up to $h=h_c$, at 
which a first-order phase transition occurs. Therefore, we fix $h/\Delta_0=0.3$ which satisfies $h<h_c$ for 
all symmetries considered. An important point in the context of $p$-wave superconductors is that the applied field $\mathbf{B}$ must be parallel to the $\mathbf{d}_\vk$-vector to probe the Pauli limiting effect. In the present paper, both $\mathbf{B}$ and $\mathbf{d}_\vk$ are assumed to lie in the plane of the thin-film superconductor. Note that Zeeman-splitting of unconventional superconductors have been accomplished experimentally; see \eg Ref.~\onlinecite{aprili} for the $d$-wave case.
\par
To illustrate the physics in a simple manner, we employ a two-dimensional calculation in the clean limit using the 
framework developed in Ref. \onlinecite{BTK}. Our calculations are done in the zero temperature limit, $T\to0$. We consider positive excitation energies of the incoming electrons from the 
normal side, which places the restriction that $h<\Delta_\vk$ in the superconductor. For the $s$-wave and chiral $p$-wave 
symmetry, this translates to $h<\Delta_0$ which is satisfied for our choice $h/\Delta_0=0.3$. For the $p_x$-wave symmetry, 
we must have $h<\Delta\cos\theta$. Physically, the contribution to the current will be strongest for normal incidence of 
the quasiparticles with respect to the tunneling barrier, such that we may, to a good approximation, introduce an upper 
cut-off $\theta_c$ in the angular integration of the current. For $\theta_c \simeq 75^\circ$, $h/\Delta_\vk<0.3$ is 
satisfied for all angles of incidence in the $p_x$-wave case. For the $s$-wave and chiral $p$-wave case, we use 
$\theta_c=\pi/2$.
\par
In this approach, the expression for the spin resolved tunneling current may be written as
\begin{align}
I_\alpha (eV) &= I_0\int^{\theta_c}_{-\theta_c} 
\int^\infty_{-\infty} \text{d}\theta\text{d}\varepsilon \cos\theta[f(\varepsilon-eV)-f(\varepsilon)]\notag\\
&\times[ 1 + |r^\alpha_A(\varepsilon,\theta)|^2 - |r^\alpha_N(\varepsilon,\theta)|^2]
\end{align}
with $\alpha=\uparrow, \downarrow$. The scattering coefficients $\{r_A,r_N\}$ may be obtained by exploiting the boundary conditions of the quasiparticle 
wavefunctions at the normal metal/superconductor (N/S) interface. Along the lines of Ref. \cite{tanaka_95}, one finds that
\begin{align}
r_a^\alpha &= \frac{4u_-^\alpha v_+^\alpha\e{-\i\gamma_+}}{u_+^\alpha u_-^\alpha (4-Z_\theta^2) 
+ Z_\theta^2v_+^\alpha v_-^\alpha \e{\i(\gamma_--\gamma_+)}},\notag\\
r_n^\alpha &= -1+\frac{2[u_+^\alpha u_-^\alpha (2+Z_\theta) 
- Z_\theta v_+^\alpha v_-^\alpha \e{\i(\gamma_--\gamma_+)}]}{u_+^\alpha u_-^\alpha (4-Z_\theta^2) 
+ Z_\theta^2v_+^\alpha v_-^\alpha \e{\i(\gamma_--\gamma_+)}},
\end{align}
with the definition $Z_\theta = Z_0/(\i\cos\theta)$ where $Z_0 = 2mV_0/k_F$ is a measure of the strength of the scattering potential 
at the interface. We have introduced the coherence factors $u_\pm^\alpha  = u^\alpha(\varepsilon, \theta_\pm)$ and 
$v_\pm = v^\alpha(\varepsilon, \theta_\pm)$ with $\theta_+ = \theta,$ $\theta_- = \pi-\theta$, and 
\begin{align}
[u^\alpha(\varepsilon, \theta_\pm)]^2 = \Bigg[\frac{1}{2}+ \frac{\sqrt{(\varepsilon+\alpha h)^2 
- |\Delta(\varepsilon,\theta_\pm)|^2}}{2(\varepsilon+\alpha h)}\Bigg]^{1/2},
\end{align}
with $[v^\alpha(\varepsilon, \theta_\pm)]^2= 1 - [u^\alpha(\varepsilon, \theta_\pm)]^2$. The phase of the superconducting gap is contained 
in the factor 
\begin{align}
\e{\i\gamma_\pm} = \e{\i\gamma(\theta_\pm)} = \Delta(\varepsilon,\theta_\pm)/|\Delta(\varepsilon,\theta_\pm)|.
\end{align}
Also, we 
have made use of the quasiclassical approximation $\varepsilon_F \gg (\varepsilon, \Delta)$ such that the wavevector $k_\theta = k_F\cos\theta$ 
is the same on the normal and superconducting side. Moreover, the spin-polarization of the current is given by
\begin{align}
P=(I_\uparrow - I_\downarrow)/(I_\uparrow+I_\downarrow).
\end{align}
In what follows, we will compare an intermediate transparency barrier to a low transparency barrier, as these two cases are the most 
realistic scenarios experimentally. Note that for all symmetry states considered here, the spin-part of the Cooper pair wave-function 
has $S_z=0$. As we 
shall see, the spin-polarization is nevertheless strongly affected by the differing orbital symmetries of the superconducting states.

\section{Results}
\begin{figure}[h!]
\centering
\resizebox{0.49\textwidth}{!}{
\includegraphics{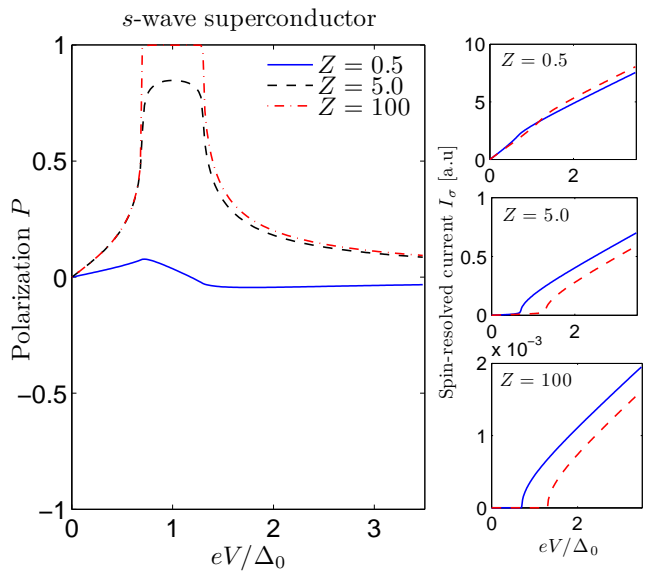}}
\caption{(Color online) Plot of the spin-polarization of the tunneling current for an $s$-wave symmetry. In the right panels, the 
full-drawn (dashed) line corresponds to majority (minority) spin.}
\label{fig:swave}
\end{figure}

We begin by commenting briefly on the $s$-wave symmetry, which was recently treated in Ref. \cite{taddei} (see Fig. \ref{fig:swave}). By increasing the barrier strength $Z$, a fully spin-polarized current is generated in the regime $|eV-\Delta_0|\leq h$. This may be 
understood by the fact that the spin-$\uparrow$ and spin-$\downarrow$ currents begin to flow at different voltages, as experimentally verified in Ref. \cite{meservey}.
\begin{figure}[h!]
\centering
\resizebox{0.48\textwidth}{!}{
\includegraphics{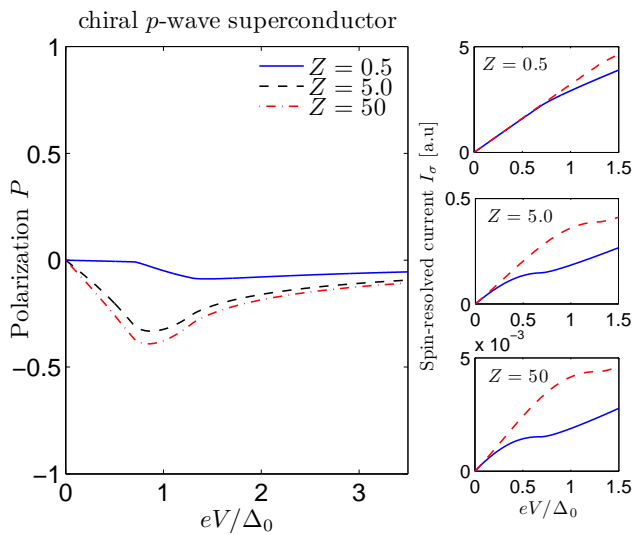}}
\caption{(Color online) Plot of the spin-polarization of the tunneling current for a chiral $p$-wave symmetry. In the right panels, 
the full-drawn (dashed) line corresponds to majority (minority) spin. }
\label{fig:chiral}
\end{figure}
\par
\begin{figure}[h!]
\centering
\resizebox{0.49\textwidth}{!}{
\includegraphics{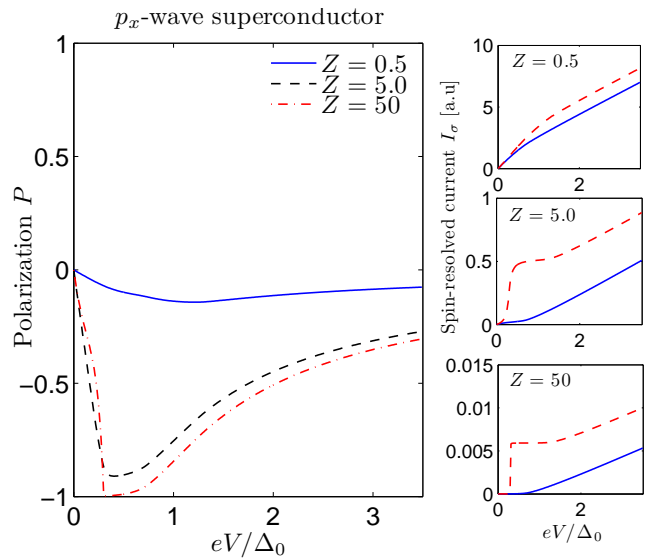}}
\caption{(Color online) Plot of the spin-polarization of the tunneling current for a $p_x$-wave symmetry. In the right panels, the 
full-drawn (dashed) line corresponds to majority (minority) spin.}
\label{fig:pwave}
\end{figure}

Next, we consider a chiral $p$-wave symmetry in Fig. \ref{fig:chiral}, believed to be realized in Sr$_2$RuO$_4$.\cite{maeno} For an 
intermediate barrier transparency ($Z=0.5$), the polarization is similar to the $s$-wave case. Increasing the barrier strength to 
$Z=5.0$ and $Z=50$, however, the polarization actually becomes dominated by the minority spin carriers. Before explaining the physics behind this unusual behaviour, we consider the $p_x$-wave symmetry below. The result is shown in Fig. \ref{fig:pwave}, in which case there is a formation of 
zero-energy Andreev bound states near the interface \cite{hu, tanaka_95}. 
Such a pairing symmetry might be realized in the heavy fermion compound UGe$_2$ \cite{Harada_2007}. Zero-energy 
states are also allowed to form in the chiral 
$p$-wave case, but only for angles of incidence $\theta=0$. In the $p_x$-wave case, the effect of zero-energy states may therefore be expected 
to be much more prononunced since all quasiparticle trajectories contribute to the formation of these surface states. Comparing Figs. \ref{fig:chiral} and \ref{fig:pwave}, one may immediately infer that this is so. Qualitatively, they are very similar, but the polarization is in general stronger in the $p_x$-wave case for a given value of $Z$. The most striking 
aspect of the polarization for both the chiral $p$-wave and $p_x$-wave symmetry is that it is exclusively negative, which means that the minority spin carriers dominate 
the transport for positive voltage. In fact, the tendency of the polarization with increasing barrier strength is \textit{opposite} 
to the $s$-wave case: a fully spin-polarized current consisting of minority spin carriers is obtained in the tunneling limit. It is 
quite remarkable that the polarization actually favors the minority spin-carriers even though the majority spin carriers benefit 
energetically from the presence of the exchange field. In order to understand this interesting behaviour, recall that in the absence 
of a magnetic field there will be an immediate onset of electrical current at zero bias due to the formation of zero-energy states 
at the interface \cite{hu, tanaka_95}. In the present case, the exchange field will split the spin-bands such that the onset of the 
minority-spin current occurs at $eV=h$ instead of $eV=0$. The majority-spin current, on the other hand, will experience the sharp 
onset of current flow at $eV=-h$. Therefore, if the symmetry of the superconducting order parameter is such that it may accommodate 
zero-energy surface states, the tendency of the polarization will be towards being \textit{negative} for positive voltages. Note 
that the polarization goes to zero at $eV=0$ for all symmetry states considered here. The same tendency was seen for the chiral 
$p$-wave case in Fig. \ref{fig:chiral}.
\par
\begin{figure}[h!]
\centering
\resizebox{0.49\textwidth}{!}{
\includegraphics{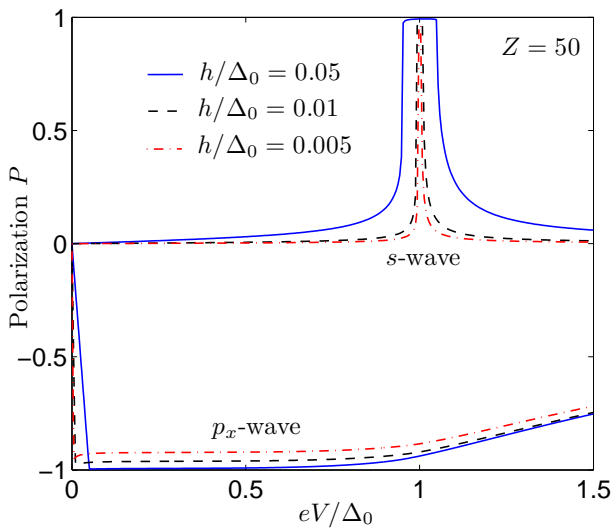}}
\caption{(Color online) Comparing the $s$-wave and the $p_x$-wave symmetry for different exchange fields at $Z=50$.}
\label{fig:swaveVSpwave}
\end{figure}

Tuning the strength of the applied magnetic field permits full control over the induced exchange energy $h$. For very weak exchange energies $h/\Delta_0 \ll1$, a major advantadge of using superconductors with zero-energy states to obtain strongy polarized currents becomes evident. In Fig. \ref{fig:swaveVSpwave}, we compare the $s$-wave and $p_x$-wave symmetry against each other for $Z=50$. As seen, the width of the region of full spin-polarization in the $s$-wave case is $2h$, which becomes very narrow for decreasing $h$. In stark contrast, the current remains almost fully spin-polarized in the $p_x$-wave case over virtually the entire subgap energy regime. Also note that for $h/\Delta_0\ll1$, the effective angular integration range includes the entire half-circle $\theta\in[\pi/2,-\pi/2]$ even for the $p_x$-wave symmetry.

\section{Discussion}
Let us now discuss some aspects of our model. When the magnetic field splits the spin-bands in a spin-singlet superconductor, the 
Cooper pair gains a finite center-of-mass momentum $q=2h/v_F$. This leads to the possibility of a spatially modulated superconducting 
order, known as the Fulde-Ferrel-Larkin-Ovchinnikov (FFLO) phase \cite{fflo}, however this phase has not been unambiguously observed to date 
\cite{Matsuda}. Ref.\cite{tanaka_fflo} recently studied the tunneling conductance for the FFLO-state. Here, we are considering homogeneous coexistence of the magnetic and superconducting order which occurs as long as 
one stays well below the Clogston limit \cite{clogston} ($h/\Delta_0 < 1/\sqrt{2}$ for $s$-wave superconductors).
It is also important to emphasize that we here consider electrical transport parallel to the film of the superconductor, which places restrictions on the resistances of the interfaces of our setup in Fig. \ref{fig:model}. Specifically, the bias electrode should be connected to the edge of the superconducting film as opposed to the normal situation where the electrode is attached on top of the film. We also underline that we have focused mainly on the \textit{tunneling limit} $(Z\gg1)$, which is experimentally most feasible, but also contrasted this regime with a higher barrier transparency $Z=0.5$. The splitting of the zero-energy peak originating with the surface states in the $p$-wave case is less pronounced for low values of $Z$, and thus yields a quantitatively reduced polarization compared to $Z\gg1$, although the qualitative tendency is the same.
\par
Although we have focused on the zero temperature limit in this work, our results should not be affected by any finite temperature effects as long as $T$ is not too close to $T_c$, i.e. $T/T_c \simeq 1$. As shown in Fig. 4 of Ref.~\onlinecite{taddei}, the finite temperature merely amounts to a smearing of the polarization curves for $T/T_c\ll 1$, while a reduction of the polarization properties is observed when $T$ becomes similar to $T_c$ in magnitude. Note that the existence of zero-energy surface states at the interface of a normal metal and unconventional superconductor does not depend on the temperature as long as $T<T_c$, such that the mechanism here proposed for generation of a strongly spin-polarized current should be a robust feature also at finite temperatures.

\section{Summary}
In summary, we have investigated the tunneling current in a thin-film normal/superconductor junction in the presence of an external, 
in-plane magnetic field. We have considered an $s$-wave, chiral $p$-wave, and $p_x$-wave symmetry for the superconductor. Remarkably, 
we find that even though the spin-structure of the superconducting state is similar in all three cases ($S_z=0$, opposite-spin pairing), 
the spin-polarization of the tunneling current is strongly modified by the orbital symmetry of the superconducting state. We find that 
the spin-polarization may be substantial for tunneling barriers and that there is an unusual interplay between zero-energy states and 
the magnetic field that may result in a fully spin-polarized current for minority spin carriers. We have studied 
the generation and manipulation of a strongly (and even possibly \textit{fully}) spin-polarized current by applying  of a 
weak static in-plane magnetic field to an N/S-junction, and then varying a bias voltage. Clearly, the main challenge in 
spintronics today is obtaining a clear-cut experimental technique of measuring the spin-polarization of an electrical 
current. Our findings suggest an alternative approach to obtain fully spin-polarized currents which does not rely 
on strong magnetic fields or half-metallic compounds. We have pointed to two compounds, namely Sr$_2$RuO$_4$ and 
UGe$_2$, as promising spin-triplet superconducting materials where these phenomenon should  be particularly pronounced. 

\acknowledgments
J.L. and A.S. were supported by the Research Council of Norway, Grants No. 158518/431 and No. 158547/431 (NANOMAT), and 
Grant No. 167498/V30 (STORFORSK). T. Y. acknowledges support by JSPS.


\begin{thebibliography}{99}

\bibitem{Awschalom} {\it Semiconductor Spintronics and Quantum Computation}, edited by D. D. Awschalom, D. Loss, and N. Samarth 
(Springer, Berlin, 2002).

\bibitem{Maekawa} {\it Spin Dependent Transport in Magnetic Nanostructures}, edited by S. Maekawa and T. Shinjo (Taylor and Francis, 
London, 2002).

\bibitem{zutic} I. Zutic, J. Fabian, and S. Das Sarma, Rev. Mod. Phys. \textbf{76}, 323 (2004).

\bibitem{Tserkovnyak} Y. Tserkovnyak, A. Brataas, G. E. W. Bauer, and B. I. Halperin, Rev. Mod. Phys. \textbf{77}, 1375 (2005).

\bibitem{accumulate} J. Martinek \etal, Phys. Rev. B \textbf{66}, 014402 (2002).
\bibitem{Kato} Y. K. Kato, R. C. Myers, A. C. Gossard, and D. D. Awschalom, Science \textbf{306}, 1910 (2004).

\bibitem{Wunderlich} J. Wunderlich, B. Kaestner, J. Sinova, and T. Jungwirth, Phys. Rev. Lett. \textbf{94}, 047204 (2005).

\bibitem{detect} Q.-F. Sun, H, Guo, and J, Wang, Phys. Rev. B \textbf{69}, 054409 (2004).

\bibitem{brataas} A. Brataas and Y. Tserkovnyak, Phys. Rev. Lett. \textbf{93}, 087201 (2004).

\bibitem{eremin} I. Eremin, F. S. Nogueira, and R.-J. Tarento, Phys. Rev. B \textbf{73}, 054507 (2006).

\bibitem{linder} M. S. Gr{\o}nsleth \etal, Phys. Rev. Lett. \textbf{97}, 147002 (2006).

\bibitem{eschrig} T. Champel, T. L{\"o}fwander, and M. Eschrig, Phys. Rev. Lett. \textbf{100}, 077003 (2008).

\bibitem{dani} D. Huertas-Hernando, Yu. V. Nazarov, and W. Belzig, Phys. Rev. Lett. \textbf{88}, 047003 (2002).

\bibitem{taddei} F. Giazotto and F. Taddei, Phys. Rev. B \textbf{77}, 132501 (2008).

\bibitem{maeno} A. P. Mackenzie and Y. Maeno, Rev. Mod. Phys. {\bf 75}, 657 (2003).

\bibitem{Tou_1998}  H. Tou {\it et al}, Phys. Rev. Lett., {\bf 80}, 3129 (1998).

\bibitem{Lee_2002} I. J. Lee {\it et al}, Phys. Rev. Lett., {\bf 88}, 017004 (2002).

\bibitem{Harada_2007} A. Harada {\it et al}, Phys. Rev. B {\bf 75}, 140502 (2007).

\bibitem{meservey} R. Meservey and P. M. Tedrow, Phys. Rep. 
{\bf 238}, 173 (1994).

\bibitem{clogston} A. M. Clogston, Phys. Rev. Lett. \textbf{9}, 266 (1962);
B. S. Chandrasekhar, Appl. Phys. Lett. \textbf{1}, 7 (1962).

\bibitem{aprili} M. Covington, M. Aprili, E. Paraoanu, L. H. Greene, F. Xu, J. Zhu, and C. A. Mirkin, Phys. Rev. Lett. \textbf{79}, 277 (1997)

\bibitem{BTK} G. E. Blonder, M. Tinkham, and T. M. Klapwijk,
Phys. Rev. B {\bf 25}, 4515 (1982). 

\bibitem{tanaka_95} Y. Tanaka and S. Kashiwaya, Phys. 
Rev. Lett. {\bf 74}, 3451 (1995).

\bibitem{hu} C.-R. Hu, Phys. Rev. Lett. {\bf 72}, 1526 (1994).

\bibitem{fflo} P. Fulde and R. A. Ferrell, Phys. Rev. \textbf{135}, A550 (1965); 
A. I. Larkin and Yu. N. Ovchinnikov, Sov. Phys. JETP \textbf{20}, 762 (1965).

\bibitem{Matsuda} Y. Matsuda and H. Shimahara, 
 J. Phys. Soc. Jpn. {\bf 76}, 051005 (2007).
 
\bibitem{tanaka_fflo} Y. Tanaka \etal, Phys. Rev. Lett. \textbf{98}, 077001 (2007)


\end{thebibliography}
\end{document}